\documentclass[
preprint,
showpacs,
showkeys,
amsmath,amssymb,
aps,
prb,
]{revtex4-1}

\usepackage{graphicx}
\usepackage{dcolumn}
\usepackage{bm}

\begin{document}


\title{Lattice dynamics model calculation of Kapitza conductance at solid-fluid interfaces}

\author{Sanghamitra Neogi}
\altaffiliation[Present address: ]{Department of Aerospace Engineering Sciences, University of Colorado Boulder, Boulder, Colorado 80309, USA}
\email{sanghamitra.neogi@colorado.edu}
\author{Gerald D. Mahan}
\affiliation{Department of Physics, \\Pennsylvania State University, \\University Park, PA 16802, USA}

\date{\today}

\begin{abstract}
Existing theoretical models of the interfacial thermal conductance, i.e., Kapitza conductance, of insulating solid-fluid interfaces only consider bulk properties, e.g., acoustic mismatch model and diffuse mismatch model. In this work, we propose a classical lattice dynamical model calculation of the Kapitza conductance, thereby incorporating interfacial structural details. In our model, we assume that heat is mostly carried by phonons in the solid, and that sound waves carry diffusive heat from the interface into the bulk of the liquid, where both longitudinal and transverse sound waves are considered. Sound wave dispersion is calculated from the fluid pair distribution function, evaluated using approximate integral equation theories (i.e., Percus-Yevick, Hypernetted-chain approximation). The Kapitza conductance of the solid-fluid interface is obtained from the phonon transmission coefficient at the interface. We determine the interfacial phonon transmission coefficient by solving the coupled equations of motion for the interfacial solid and fluid atoms. As an illustrative example, we derive the Kapitza conductance of solid argon-fluid neon interface, 
with pair-wise Lennard-Jones interactions.
\end{abstract}

                              
\maketitle

\section{Introduction}

The study of thermal transport across material interfaces has gained increased interests in recent years \cite{cahill2014nanoscale}, mainly due to the ongoing developments in the field of nanotechnology. At submicron length scales, the transport of thermal energy in heterogeneous \cite{Losego, OBrien, Cheng} and nanostructured materials (e.g. superlattices \cite{Chen}, nanofluids \cite{Keblinski, PKeblinski, Wang, Huxtable}, polymer nanocomposite materials \cite{Carlberg, Wen}) is strongly influenced by thermal resistance across interfaces. The presence of a solid-fluid interface strongly impacts the heat dissipation of a nanostructure to its fluid surroundings \cite{Huxtable}. Thermal transport across a solid-fluid interface is an important issue in the thermal management of photovoltaic and fuel-cell technologies \cite{Faghri, Royne} as well. When heat is conducted from one material to another, a temperature discontinuity, $\Delta T$, arises at the interface between the two materials. For small heat flow across the interface, the discontinuity is proportional to the heat flow, $J_{Q}$:
\begin{eqnarray}
J_{Q} & = & G \Delta T.\label{eq:Kapitza_conductance_definition}
\end{eqnarray}
The proportionality constant $G$ is the thermal conductance of the interface, also known as Kapitza conductance, after Kapitza \cite{Kapitza}.

Heat transport in materials can be explained in terms of the motion of the microscopic heat carriers, typically electrons and phonons. Transmission of thermal energy across an interface is hindered by the mismatch between the materials, creating resistance to the heat flow. The first theoretical formulation of the interfacial thermal conductance was presented by Khalatnikov \cite{Khalatnikov}, known as the acoustic mismatch model (AMM). AMM makes the assumption that phonon propagation in materials is governed by continuum acoustics. It considers the reflection and transmission of classical heat waves at the interface, though its continuum character prevents it from incorporating structure of the interface on an atomic scale. 
In the diffuse mismatch model (DMM), proposed by Schwartz \cite{Schwartz1}, the interfacial thermal conductance is calculated from the mismatch of the bulk vibrational density of states of the two systems across the interface. These two widely used models, fail to provide a reasonable estimate of the thermal conductance of solid-fluid interfaces, which can be used for comparison with molecular dynamics results \cite{Reddy}. As an improvement to AMM and DMM, a lattice dynamical model was proposed by Young and Maris \cite{YM} and later extended by Pettersson and Mahan \cite{PM} (YMPM) to obtain the Kapitza conductance of the interface between two solids. YMPM has proven effective for solid-solid interfaces without defects. We note, however, the comparatively slow progress of theoretical formulations for solid-{\em fluid} interfaces. Most recent efforts were provided by molecular dynamics simulations \cite{Ge, Ikeshoji, Xue1, Xue2, Sachdeva, Sarkar, Ywang, Issa, Barrat, Hu, Liang, Shenogina, Goicochea, Acharya, Landry}. 

Here, we propose an analytical model to calculate the interfacial thermal conductance of insulating solid-fluid interfaces, by considering the coupling between the phonons in the solid and sound waves in the fluid, at the interface. This model is based on classical lattice dynamics calculations of atom vibrations at the interface and is akin to the YMPM lattice dynamical model of Kapitza conductance of solid-solid interfaces. We discuss the transmission of thermal energy between phonons in the solid and sound waves in the fluid. However, a complete theory should also include fluid convection. To correctly estimate the conductance, all modes, longitudinal and transverse, on both sides of the interface were included. Transverse sound waves are known to exist in fluids at larger values of wave vector \cite{Hansen, March, Chung,Schaich, Bron}, and they can carry heat away from the interface. As an illustrative example, we calculate the interfacial thermal conductance between two Lennard-Jones systems, namely the interface between solid argon and fluid neon. Some important aspects of this model have been discussed in a previous paper by Mahan \cite{Mahan_kapitza} and we will refer to it frequently throughout this article.

The article is structured as follows: In Sec. \ref{general_theory}  we review the basics of thermal transport across interfaces. We discuss the methods to determine the vibrational properties of the two bulk systems in Sec. \ref{bulk_systems}. In Sec. \ref{interface} we describe the model\rq{}s inclusion of the interfacial properties --- the structure of the fluid near the interface and the motion of the interfacial atoms. The Subsection \ref{mode_matching} contains a discussion about the matching of the vibrational excitations of the two systems. In Section \ref{KC} we examine the evaluation of the Kapitza conductance using the phonon transmission coefficients at the interface. We illustrate the applicability of our model by calculating the Kapitza conductance of the solid argon-fluid neon interface in Section \ref{example}.

\section{General theory}
\label{general_theory}

In a solid, the heat flow per unit area per unit time along a unit normal vector $\hat{z}$ can be expressed as:
\begin{equation}
J_{Q}=\sum_{\lambda}\int\frac{{\rm d}^{3}q}{(2\pi)^{3}}\hbar\omega_{\lambda}({\bf q})\hat{z}\cdot{\bf v}_{\lambda}({\bf q})n_{B}[\omega_{\lambda}({\bf q}),T],\label{eq:heat_flow_solid}
\end{equation}
where ${\bf q}$ and $\lambda$ are the wave vector and the polarization of a phonon with energy $\hbar\omega_{\lambda}({\bf q})$, respectively. $\hat{z}\cdot{\bf v}_{\lambda}({\bf q})$ is the projection of the phonon group velocity along the normal direction $\hat{z}$ and $n_{B}[\omega_{\lambda}({\bf q}),T]$ is the Bose-Einstein occupation factor of phonons at temperature $T$. The integral over the wave vector is limited to values for which $\hat{z}\cdot{\bf v}_{\lambda}({\bf q}) \geq 0$. The Kapitza conductance of an interface $G(T)$ at temperature $T$ is the ratio of the net heat flow across the interface per unit area per unit time to the temperature difference across the interface (Eq. \ref{eq:Kapitza_conductance_definition}). If we consider an interface between materials A and B, the net heat flow across the interface per unit area per unit time can be obtained from
\begin{align}
J_{QT}&=\sum_{\lambda}^{A}\int\frac{{\rm d}^{3}q}{(2\pi)^{3}}\hbar\omega_{\lambda}^{(A)}({\bf q})\hat{z}\cdot{\bf v}^{(A)}_{\lambda}({\bf q})n_{B}[\omega^{(A)}_{\lambda}({\bf q}),T^{(A)}]\mathcal{T}_{\lambda}({\bf q}) \nonumber\\ & - \sum_{\lambda}^{B}\int\frac{{\rm d}^{3}q}{(2\pi)^{3}}\hbar\omega^{(B)}_{\lambda}({\bf q})\hat{z}\cdot{\bf v}^{(B)}_{\lambda}({\bf q})n_{B}[\omega^{(B)}_{\lambda}({\bf q}),T^{(B)}]\mathcal{T}_{\lambda}({\bf q}),\label{eq:transmitted_flux}
\end{align}
where $\mathcal{T}_{\lambda}({\bf q})$ is the phonon transmission coefficient across the interface, $\mathcal{T}^{A\rightarrow B}_{\lambda}({\bf q}) $$= \mathcal{T}^{B\rightarrow A}_{\lambda}({\bf q}) $$= \mathcal{T}_{\lambda}({\bf q})$. $T^{(A)}$ and $T^{(B)}$ are the temperatures at the two sides of the interface, and the negative sign in the second equation appears because $\hat{z}$ is defined in the direction A $\rightarrow$ B. Following the procedure described in Reference \citenum{YM}, one can obtain the following expression of the Kapitza conductance of the A-B interface:
\begin{align}
G(T)&=\frac{\partial}{\partial T}\sum_{\lambda}^{A}\int\frac{{\rm d}^{3}q}{(2\pi)^{3}}\hbar\omega_{\lambda}^{(A)}({\bf q})\hat{z}\cdot{\bf v}^{(A)}_{\lambda}({\bf q})n_{B}[\omega^{(A)}_{\lambda}({\bf q}),T]\mathcal{T}_{\lambda}({\bf q}) \nonumber\\ 
& = -\frac{\partial}{\partial T} \sum_{\lambda}^{B}\int\frac{{\rm d}^{3}q}{(2\pi)^{3}}\hbar\omega^{(B)}_{\lambda}({\bf q})\hat{z}\cdot{\bf v}^{(B)}_{\lambda}({\bf q})n_{B}[\omega^{(B)}_{\lambda}({\bf q}),T]\mathcal{T}_{\lambda}({\bf q}).\label{eq:KC}
\end{align}
We can expand the Bose-Einstein occupation factor in powers of $\frac{\hbar \omega}{k_B T}$. At high temperatures, keeping only the first order term the Kapitza conductance can be found from the following expression:
\begin{equation}
G = k_{B}\sum_{\lambda}\int\frac{{\rm d}^{3}q}{(2\pi)^{3}}\hat{z}\cdot{\bf v}_{\lambda}({\bf q})\mathcal{T}_{\lambda}({\bf q}),\label{eq:final_KR}
\end{equation}
where $k_{B}$ is the Boltzmann constant. The integral is evaluated only for phonons propagating towards the interface, so that $\hat{z}\cdot{\bf v}_{\lambda}({\bf q}) \geq 0$. 

\section{Description of the two bulk systems}
\label{bulk_systems}

\begin{figure}
\begin{centering}
\includegraphics[scale=0.4]{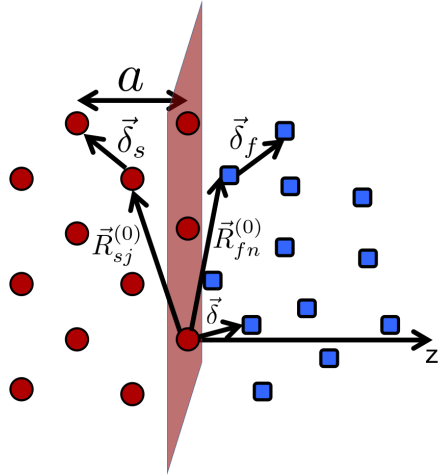}
\par\end{centering}
\caption{Illustration of a solid-fluid interface. Solid atoms are on the left and marked by circles, while fluids are on the right and marked by squares. The interface lies in the $x-y$ plane and is perpendicular to the $z$ axis in the figure. The position of one of the solid atoms at the interface is chosen to be the origin of the coordinate system. The lattice constant of the solid is denoted by $a$. Arrows show the position vectors of the solid and the fluid atoms and also the neighboring atoms. The different vectors shown in the figure are described in the text. \label{fig:Cartoon_solid_fluid}}
\end{figure}

We discuss heat transport across the interface between an insulating solid and an insulating fluid. In particular, we study the interface between a solid with a face-centered cubic (FCC) lattice and a classical fluid with atoms interacting via the Lennard-Jones (LJ) potential. A schematic sketch of the interface is shown in Fig. \ref{fig:Cartoon_solid_fluid}, where the interface is marked by the plane perpendicular to the $z$ axis.
Solid atoms are on the left of the interface and marked with circles, while fluid atoms are to the right of the interface and marked with squares. $\vec{R}_{sj}^{(0)}$ denotes the equilibrium position of the $j^{\text{th}}$ solid atom and $\vec{R}_{fn}^{(0)}$ denotes the equilibrium position of the $n^{\text{th}}$ fluid atom. The distances between the neighboring atoms in the solid are represented by the vector ${\bf \delta}_{s}$ and that in the fluid by ${\bf \delta}_{f}$. The distance between a solid and a neighboring fluid atom near the interface is represented with ${\bf \delta}$. These vector notations are used throughout. 

\subsection{Phonons in the solid}

\begin{figure}
\begin{centering}
\includegraphics[scale=0.4]{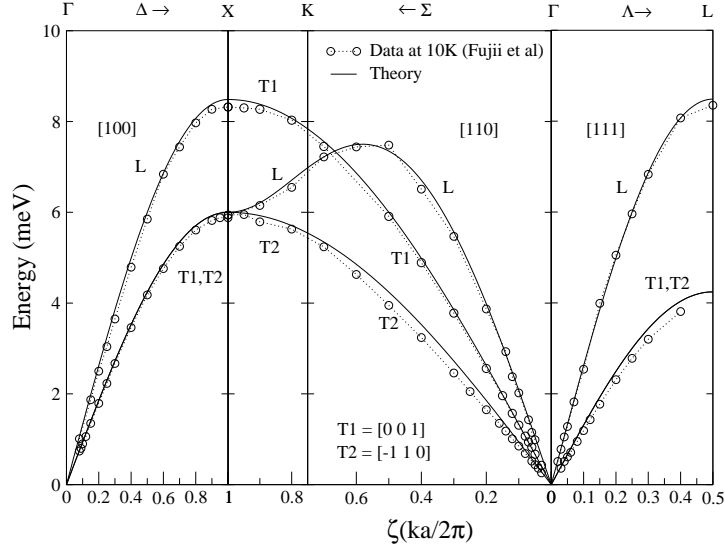}
\par\end{centering}
\caption{Phonon dispersion in solid argon along the different symmetry directions of the FCC lattice, obtained using harmonic lattice dynamics. The theoretical values are shown by the solid lines and the experimental data points are marked by the open circles connected with dashed lines. Experimental data taken from Fujii et al \cite{Fujii}. Only nearest neighbor interaction is considered to obtain the theoretical dispersion data.\label{fig:solid_phonons}}
\end{figure}

We obtain the phonon dispersion in the FCC solid using harmonic lattice dynamics, the procedure being described in the Appendix. As an illustrative example, we compute the phonon dispersion in FCC solid argon. The value of the lattice parameter of argon we use in our calculation is 5.31 \AA. To obtain the nearest and the next-nearest neighbor spring constants $K_{1}$ and $K_{2}$, respectively, we compare the analytical expressions of $\omega$ at the different symmetry points of the FCC lattice to the experimentally measured values shown in Fig. \ref{fig:solid_phonons}. The value of $K_{2}$ is found to be almost negligible, while $\hbar\sqrt{K_{1}\slash M_{s}} \approx 3.00$ meV. 
In Fig. \ref{fig:solid_phonons}, we show the theoretically obtained values of phonon dispersion in solid argon (i.e., using $K_1$) in comparison with experiments. The agreement indicates that nearest-neighbor interaction is sufficient to accurately reproduce the complete phonon dispersion in solid argon at low temperatures. Our calculations of phonon dispersion in other noble gas solids (Kr and Xe) also confirm that nearest-neighbor interaction alone is sufficient to reproduce the complete phonon dispersion at low temperatures, consistent with experimental results (data not shown for brevity). To the best of our knowledge, this result has not yet been reported in the literature. This result reduces the computational effort required in this endeavour. In the following, we thus only consider nearest-neighbor interactions in the solid. Also, we keep $K_1 = 3.00$ meV for argon in all following calculations, with the assumption that this value does not change significantly with temperature.
 
\subsection{Sound wave excitations in the fluid}
\label{sound_waves_bulk}

We consider sound waves as the heat carriers in classical fluids. The dispersion of sound waves in classical fluids in equilibrium can be determined by the general expression \cite{Hansen, Zwanzig1, Zwanzig2}
\begin{equation}
\omega_{{\bf q}\lambda}^{2}=\frac{k_{B}T}{M_{f}}(1+2\hat{\epsilon}_{{\bf q}\lambda}\cdot\hat{\epsilon}_{{\bf q}1})q^{2}+\frac{n}{M_{f}}\int {\rm d}^3{\bf r}g({\bf r})[1-\exp({\bf q}\cdot{\bf r})](\hat{\epsilon}_{{\bf q}\lambda}\cdot{\bf \nabla})^{2}V_{ff}({\bf r}),\label{eq:frequency spectrum}
\end{equation}
where $\omega_{{\bf q}\lambda}$ is the frequency of the sound wave in the fluid, $n$ is the density of the fluid and $M_{f}$ is the mass of a fluid atom. $V_{ff}({\bf r})$ is the interaction potential between the fluid atoms and $g({\bf r})$ is the pair distribution function of the fluid atoms in the bulk. $\hat{\epsilon}_{{\bf q}\lambda}(\lambda=1,2,3)$ represent the three orthonormal polarization vectors of the sound waves in the fluid, where the longitudinal one is denoted by
\begin{equation}
\hat{\epsilon}_{{\bf q}1}=\frac{{\bf q}}{|{\bf q}|}.\label{eq:longitudinal polarization}
\end{equation}
The first term on the R.H.S of Eq. (\ref{eq:frequency spectrum}) is $ 3 k_{B}Tq^{2}/M_{f}$ for the longitudinal mode and $k_{B}Tq^{2}/M_{f}$ for the two transverse modes. The second term can be written explicitly in terms of its components
\begin{align}
&\frac{n}{M_f}\sum_{i,j=1}^{3}\epsilon_{q_{i}\lambda}\epsilon_{q_{j}\lambda}\int {\rm d}^3{\bf r}g({\bf r})[1-\exp({\bf q}\cdot{\bf r})]\nonumber \\ 
&\times\left[\frac{\delta_{ij}}{r}\frac{{\rm d}V_{ff}(r)}{{\rm d}r}+\frac{r_ir_j}{r^2}\left(\frac{{\rm d}^2V_{ff}(r)}{{\rm d}r^2}-\frac{1}{r}\frac{{\rm d}V_{ff}(r)}{{\rm d}r}\right)\right].\label{eq:frequency_integral_term}
\end{align}
We assume that the interaction between the fluid atoms can be described by the Lennard-Jones (LJ) potential,
\begin{equation}
V(r)=4\epsilon\left(\frac{\sigma^{12}}{r^{12}}-\frac{\sigma^{6}}{r^{6}}\right), \label{eq:LJ_potential}
\end{equation}
where $r$ is the interatomic distance, $\sigma$ is the interaction range and $\epsilon$ is the well-depth of the LJ potential. We assume a harmonic approximation for the fluid-fluid interaction potential and use the approximate potential form given in the Appendix (Eq. (\ref{eq:fluid_potential_Taylor_expansion})). We are not aware of previous studies using such a harmonic approximation along with a Lennard-Jones interaction potential to obtain sound waves in classical fluids --- a novel approach, in particular in the context of thermal transport. For the LJ potential (Eq. (\ref{eq:LJ_potential})), the derivatives of the potential $V_{ff}(r)$ can be written as
\begin{eqnarray}
A_{ff}(r) & = & \frac{1}{r}\frac{{\rm d}V_{ff}(r)}{{\rm d}r} = -\frac{24\epsilon}{\sigma^{2}}\left(\frac{2\sigma^{12}}{r^{14}}-\frac{\sigma^{8}}{r^{8}}\right) ,\text{ and, }\label{eq:A}\\
B_{ff}(r) & = & \frac{{\rm d}^2V_{ff}(r)}{{\rm d}r^2}-\frac{1}{r}\frac{{\rm d}V_{ff}(r)}{{\rm d}r} = \frac{96\epsilon}{\sigma^{2}}\left(\frac{7\sigma^{12}}{r^{14}}-\frac{2\sigma^{8}}{r^{8}}\right).\label{eq:B}
\end{eqnarray}
To solve the integral in Eq. (\ref{eq:frequency_integral_term}), we choose a coordinate system in which ${\bf q}$ lies along the $z$-axis of the coordinate system without any loss of generality. The wavevectors in the old and new systems are denoted by ${\bf q} = (q_1, q_2, q_3)$ and ${\bf q}^{\prime} = (0, 0, q^{\prime}_3)$, respectively. They are related by $\sum_{i=1}^3 q^2 = q^{\prime2}_{3}$. In this new coordinate system, ${\bf r}$ makes an angle $\theta$ with ${\bf q}^{\prime}$ and is allowed to span all space $(r = [0, \infty], \theta = [0, \pi], \phi = [0, 2\pi])$. Evaluating the integrals in spherical polar coordinates, we obtain the following expression of the dispersion of sound waves in fluid:
\begin{eqnarray}
&&\omega_{{\bf q}^{\prime}\lambda}^{2} = (1+2\hat{\epsilon}_{{\bf q}^{\prime}\lambda}\cdot\hat{\epsilon}_{{\bf q}^{\prime}1})\omega_{0}^{2} \nonumber \\
&&+\frac{4\pi n}{M_{f}}\int r^{2}{\rm d}rg(r)\bigg \{ A_{ff}(r)\left(1-j_{0}(q^{\prime}r)\right) + B_{ff}(r)\left(\frac{1}{3}-\frac{j_{1}(q^{\prime}r)}{q^{\prime}r}\right)\label{eq:longitudinal_freq}\\
&&+ \epsilon^2_{q^{\prime}_3 \lambda} B_{ff}(r)\left(\frac{j_1(q^{\prime}r)}{q^{\prime}r}-\left(\frac{{\rm d}j_{1}(z)}{{\rm d}z}\right)_{z=q^{\prime}r}\right)\bigg \},\nonumber \\
&&\omega_{0}^{2} = q^{\prime2}\frac{k_{B}T}{M_{f}},\label{eq:zero_freq}
\end{eqnarray}
where $j_0(z)(=\sin(z)/z)$ and $j_1(z)(= \sin(z)/z^2 - \cos(z)/z)$ are spherical Bessel functions. The term $\omega_{0}^{2}$ comes from the long range density fluctuations in the fluid. The polarization vector of the longitudinal sound wave in the new coordinate system is $\hat{\epsilon}_{{\bf q}^{\prime}1} = (0, 0, 1)$. One can choose the polarization vectors of the transverse sound waves such that $\epsilon_{q^{\prime}_3 \lambda} = 0$, for $\lambda = 2,3$. Inserting the values of $\epsilon_{q^{\prime}_3\lambda}$ into Eq. (\ref{eq:longitudinal_freq}), we retrieve the fluid dispersion expressions for longitudinal and transverse sound waves given in Reference \citenum{Mahan_kapitza}. Using a coordinate transformation, we retrieve the polarization vectors in the original coordinate system: 
\begin{equation}
\hat{\epsilon}_{{\bf q}l}=\frac{(q_{x},q_{y},q_{z})}{\sqrt{q_{x}^{2}+q_{y}^{2}+q_{z}^{2}}},\hat{\epsilon}_{{\bf q}t1}=\frac{(-q_{y},q_{x},0)}{\sqrt{q_{x}^{2}+q_{y}^{2}}},\hat{\epsilon}_{{\bf q}t2}=\frac{(q_{x}q_{z},q_{y}q_{z},-(q_{x}^{2}+q_{y}^{2}))}{\sqrt{q_{x}^{2}+q_{y}^{2}+q_{z}^{2}}\sqrt{q_{x}^{2}+q_{y}^{2}}}.\label{eq:fluid-polarization}
\end{equation}
Equations (\ref{eq:longitudinal_freq}, \ref{eq:zero_freq}) and (\ref{eq:fluid-polarization}) are the main equations required to provide a complete description of sound waves in classical fluids. The essential ingredient required in Eq. (\ref{eq:longitudinal_freq}) is the fluid pair distribution function.  

We evaluate the pair distribution function in bulk fluid using integral equation theories. The approximate integral technique to obtain the particle distribution function in classical fluids was first proposed by Percus \cite{Percus62, Percusjk64} and later extended by various researchers during the 1960's and 70's \cite{Stell, DeBoer, Verlet, Levesque, Throop, Verlet2}. The pair distribution function for a uniform fluid is defined solely from the relative separation between atoms. The Percus-Yevick equation of pair distribution function of a uniform fluid is given by 
\begin{equation}
 g(r)e^{\beta V(r)} = 1+ n \int {\rm d}^3{\bf r}\left[g(|{\bf r}-{\bf r}^\prime|)-1\right]\left[1-e^{\beta V(r)}\right]g(r)\label{eq:PY_equation}
\end{equation}
and the Hypernetted-chain equation (HNC) is given by
\begin{equation}
\ln g(r) + \beta V(r) = n \int {\rm d}^3{\bf r} \left[g(|{\bf r}-{\bf r}^\prime|)-1\right]\left[g(r)-1-\ln g(r)-\beta V(r)\right].\label{eq:HNC_equation}
\end{equation}
The integral equations can be solved numerically using iterative methods to obtain the pair distribution function of the fluid. In order for these iterative methods to converge within reasonable computational time, one needs to provide judiciously chosen initial input parameters and also use effective mixing of intermediate results. 

\section{Interface between the two systems}
\label{interface}

The phonons in the solids that are incident to the interface are either reflected back to the solid or transmitted to the fluid with some reflection and transmission amplitude, respectively. Let $\omega$, $\hat{\epsilon}_{i}$ and (${\bf q}$, $q_{i}$) represent the frequency, polarization and wavevector of the incident phonon, respectively. ${\bf q}=(q_{x},\, q_{y}\,,0)$ represents the components of the wave vector parallel to the interface and $q_{i}$ is the component perpendicular to the interface. For a given value of the incident wave vector (${\bf q}$, $q_{i}$), we determine the frequency and polarization vector of the incident phonon by solving the eigenvalue problem of the dynamical matrix, as given by Eq. (\ref{eq:secular_determinant}) and Eq. (\ref{eq:solid_eigenvalue_equation}) in the Appendix, respectively. 

\subsection{Vibrational coupling between the two systems}
\label{mode_matching}

\subsubsection{Reflected waves}

The reflected waves have the same frequency $\omega$ and parallel components $q_x$ and $q_y$ as that of the incident wave, but different perpendicular components of the wavevectors $q_{s\lambda}$ and polarization vectors $\hat{\epsilon}_{s\lambda}$. Here, $s$ stands for solid and $\lambda\,(=1,2,3)$ denotes the three polarization directions. In order to obtain the perpendicular components of the wave vectors of the three reflected phonons ($q_{s\lambda}$), we need to solve Eq. (\ref{eq:secular_determinant}) using $\omega$, $q_x$, and $q_y$. Solving numerically Eq. (\ref{eq:secular_determinant}) is difficult in many cases and usually requires large amounts of computational time. We transform Eq. (\ref{eq:secular_determinant}) into an algebraic equation that can be solved easily. To carry out the transformation we define
\begin{eqnarray}
X & = & \cos\left(\frac{\theta_{x}}{2}\right),\, Y=\cos\left(\frac{\theta_{y}}{2}\right),\, Z=\exp\left(i\frac{\theta_{z}}{2}\right),\label{eq:transformed_var}\\
\text{and \,}\Omega & = & \frac{M_{s}\omega^{2}}{K_1}.\label{eq:bigomega}
\end{eqnarray}
After some algebra, Eq. (\ref{eq:secular_determinant}) can be written
in the form 
\begin{eqnarray}
&&C_{3}Z^{6}+C_{2}Z^{5}+C_{1}Z^{4}+C_{0}Z^{3}+C_{1}Z^{2}+C_{2}Z+C_{3}=0,\label{eq:z_reflected_eq}\\
&&\text{where, }\nonumber \\
&&C_{3} =X+Y,\nonumber \\
&&C_{2} =\Omega(2+3XY)-4-8XY-4(X^{2}+Y^{2})+8X^{2}Y^{2},\nonumber \\
&&C_{1} =2\Omega^{2}(X+Y)+\Omega(6XY-16)(X+Y)+27(X+Y)\nonumber \\
&& -16XY(X+Y)+4(X^{3}+Y^{3}),\nonumber \\
&&C_{0} =\Omega^{3}+\Omega^{2}(4XY-12)+\Omega(40+8(X^{2}+Y^{2})-26XY)\nonumber \\
&& -40+32XY-24(X^{2}+Y^{2})+8XY(X^{2}+Y^{2}).\nonumber 
\end{eqnarray}
For each root $Z_{l}\,(l=1,\ldots6)$, there is a corresponding root $1/Z_{l}$.  
If some of the roots have a non-zero imaginary part, we only retain the roots with positive imaginary part. The waves with real $q_{s\lambda}$ reflects off the interface and can carry the heat away from the interface. The waves with complex $q_{s\lambda}$ do not carry heat away from the interface, but are still required to satisfy the boundary conditions. With the knowledge of the wavevector (${\bf q}, q_{s\lambda}$) of the reflected waves, we can form the dynamical matrix and solve the eigenvalue equation (see Eq. (\ref{eq:solid_dynamical_matrix}) and Eq. (\ref{eq:solid_eigenvalue_equation}) in the Appendix) to obtain the polarization vectors of the reflected waves, where the eigenvalue is the incident phonon frequency $\omega$.

Once we have determined the wavevectors of the incident and reflected waves, the group velocities can be obtained in the following way. Using the transformations
given in Eq. (\ref{eq:transmission_coeff}), we can rewrite Eq. (\ref{eq:solid_eigenvalue_equation}) as 
\begin{eqnarray}
&&\Omega^{3}+A\Omega^{2}+B\Omega+C=0,\label{eq:group_velocity_eq}\\
&&\text{where, }\nonumber \\
&&A = 4(XY+YZ+ZX)-12,\nonumber \\
&&B = 8(X^2+Y^2+Z^2)+12XYZ(X+Y+Z)\nonumber \\
&&-32(XY+YZ+ZX)+36,\nonumber \\
&&C = 8(X^3Y+XY^3+X^3Z+Y^3Z+XZ^3+YZ^3)+32X^2Y^2Z^2 \nonumber \\
&&-32(X^2YZ+XY^2Z+XYZ^2) -16(X^2Y^2+X^2Z^2+Y^2Z^2) \nonumber \\
&&-16(X^2+Y^2+Z^2)+48(XY+XZ+YZ)-32.\nonumber 
\end{eqnarray}
Differentiating Eq. (\ref{eq:group_velocity_eq}) with respect to $q_z$ we obtain
\begin{align}
 v^{(s)}_z = -\sqrt{\frac{Ka^2}{4M_s\omega}} \frac{\Omega^2\frac{\partial A}{\partial Z}+\Omega \frac{\partial B}{\partial Z}+\frac{\partial C}{\partial Z}}{3\Omega^2+2A\Omega+B} \sin\left(\frac{\theta_z}{2}\right), 
\end{align}
where $v^{(s)}_z$ is the phonon group velocity in the solid along the $\hat{z}$ direction.

\subsubsection{Transmitted waves}

The transmitted waves in the fluid have the same frequency $\omega$ and parallel components of the wavevectors ${\bf q}$ as the incident phonon, though perpendicular components of the wavevectors $q_{f\lambda}$ and polarization vectors $\hat{e}_{f\lambda}$ differ. We remind the reader that $f$ stands for fluid and $\lambda=1,2,3$ represents the three polarization directions.
We can calculate the wave vectors of the transmitted sound waves in the fluid using the dispersion relation (Eq. (\ref{eq:longitudinal_freq})). Equation (\ref{eq:longitudinal_freq}) is an integral equation that is computationally expensive to invert and solve a range of $\omega$ values to consider. Instead we take the following approach to work around this problem. We approximate the function $\omega_{\lambda}({\bf q}^{\prime})$ as a polynomial in $q_{\lambda}^{\prime}$:
\begin{equation}
\omega_{\lambda}(q^{\prime})=\sum_{p=1}^{m}\alpha_{p}q_{\lambda}^{\prime p}.\label{eq:omega_polynomial}
\end{equation}
We obtain the coefficients $\alpha_{p}$'s by fitting the polynomial expansion to the dispersion data obtained by solving the integral equation for a range of $q$ values. An example of dispersion data for fluid neon is shown in Fig. \ref{fig:Ne_sound_modes}. To obtain a satisfactory fit of the functional form of $\omega_{\lambda}$ for all ranges of $q_{\lambda}^{\prime}$, we truncated the polynomial expansion at $p=6$. Use of this functional form allows us to evaluate the wavevectors of the transmitted sound waves, $q_{\lambda}^{\prime}$, for given values of $\omega_{\lambda}$ in a straightforward manner. Note that $q_{\lambda}^{\prime}$ here represents the norm of the full wavevector. As we described in Section \ref{sound_waves_bulk}, the norm of the wavevector ${\bf q}_{\lambda}^{\prime}$ is the same as the norm of the wavevector ${\bf q}_{\lambda}$ in the original system, hence, $q_{\lambda}^{\prime}=\sqrt{q_{x}^{2}+q_{y}^{2}+q_{z \lambda}^{2}}$. One can calculate the perpendicular components of the 
wavevectors of the transmitted waves in the original coordinate system $q_{z \lambda}$, using given values of the parallel components of 
the wavevectors, $q_{x}$ and $q_{y}$: $q_{z\lambda}=\sqrt{q_{\lambda}^{\prime 2}-q_{x}^{2}-q_{y}^{2}}$. The polarization vectors of the transmitted sound waves can be constructed using the method described by Eq. (\ref{eq:fluid-polarization}). The group velocities of the transmitted waves can be obtained from the numerical derivative of Eq. (\ref{eq:omega_polynomial}) with respect to $q_{z\lambda}$.

\subsection{Solid-fluid interaction at the interface}
\label{solid_fluid_interaction}

We consider that the solid atoms and the fluid atoms near the interface are interacting with a short-range potential $V_{sf}({\bf r})$. The potential is assumed to be central (i.e., $V_{sf}({\bf r}) = V_{sf}(r)$). We further assume that the phonons in the solid and the sound waves in the fluid displace the interfacial solid and the fluid atoms only by a small amount, i.e. the displacements of the interfacial atoms are much smaller than their relative separations from the neighboring atoms. Hence, we can truncate the Taylor expansion of the solid-fluid interaction potential around small displacements after second order, similar to the case of the fluid-fluid interaction potential as described in the Appendix, Eq. (\ref{eq:fluid_potential_Taylor_expansion}).

\subsubsection{Distribution of fluid atoms near the interface}
\label{fluid_structure}

Due to the interaction between the interfacial solid and the fluid atoms, the structure of the fluid is different near the interface from that in the bulk. We define the probability that there is a fluid atom at ${\bf R}_{fn}^{(0)}$ provided that there is a solid atom at ${\bf R}_{sj}^{(0)}$, by the one-particle distribution function near the interface $g_{sf}({\bf R}_{sj}^{(0)},\,{\bf R}_{fn}^{(0)})$,
\begin{equation}
g_{sf}({\bf R}_{sj}^{(0)},\,{\bf R}_{fn}^{(0)})=g\left(\left\{{\bf R}_{fn}^{(0)}-{\bf R}_{sj}^{(0)}\right\}\bigg|V_{sf}\left\{\left|{\bf R}_{fn}^{(0)}-{\bf R}_{sj}^{(0)}\right|\right\}\right).\label{eq:g_sf}
\end{equation}
The one-particle distribution function near an interface can be evaluated in the following way: according to Percus \cite{Percusjk64}, the one particle probability density $g({\bf r}|U)$ in the presence of an external potential $U$ can be written as
\begin{align}
g({\bf r}|U)e^{\beta U({\bf r})}= & g({\bf r})+\int {\rm d}^{3}{\bf r}_{1}\mathcal{F}_{2}({\bf r},{\bf r}_{1})\left(e^{-\beta U(r_{1})}-1\right)\nonumber \\
+ & \frac{1}{2}\int {\rm d}^{3}{\bf r}_{1}\int {\rm d}^{3}{\bf r}_{2}\mathcal{F}_{2}({\bf r},{\bf r}_{1},{\bf r}_{2})\left(e^{-\beta U(r_{1})}-1\right)\left(e^{-\beta U(r_{2})}-1\right)+\ldots\label{eq:one_particle_prob_density}
\end{align}
where $\mathcal{F}_{2,3}$ are the two and three particle Ursell functions \cite{Percusjk64}, respectively. Ursell functions quantify the deviation of the $n$-body probability density of the interacting system from that of the noninteracting system; for example, the two particle Ursell function is given by
\begin{equation}
\mathcal{F}_{2}({\bf r}_{1},{\bf r}_{2})=n^{(2)}({\bf r}_{1},{\bf r}_{2})-n^{(1)}({\bf r}_{1})n^{(1)}({\bf r}_{2}).\label{eq:Ursell function}
\end{equation}
Here $n^{(1),(2)}$ are the one and two particle probability densities, respectively. If we ignore three particle correlations ($\mathcal{F}_{n}=0$ for $n\geq3$), and use the relationship between the Ursell function and the pair distribution function 
\begin{equation}
\mathcal{F}_{2}({\bf r}_{1},{\bf r}_{2})=n^{2}[g({\bf r}_{1},{\bf r}_{2})-1],\label{eq:Ursell_gofr}
\end{equation}
we obtain the expression of the one-particle distribution function of a uniform fluid near an interface
\begin{align}
g({\bf r}|V_{sf})e^{\beta V_{sf}({\bf r})}\approx & 1+n\int {\rm d}^{3}{\bf r}_{1}\left[g(|{\bf r}-{\bf r}_{1}|)-1\right]\left(e^{-\beta V_{sf}(r_{1})}-1\right).\label{eq:one_particle_distr}
\end{align}
We make the simplifying assumption that the distribution of fluid particles is more affected perpendicularly to its interface compared to the parallel direction and hence, only consider the $z$-dependence of the one-particle distribution function. By partial evaluation of the integral in Eq. (\ref{eq:one_particle_distr}) in cylindrical polar coordinates, the one-particle distribution in the $z$-direction can be approximated as
\begin{align}
g(z|V_{sf})e^{\beta V_{sf}(z)}\approx & 1+2\pi n\int {\rm d}z_1 \left[g(|z-z_{1}|)-1\right] \int {\rm d}\rho_{1}\left(e^{-\beta V_{sf}(\sqrt{\rho_1^2+z_1^2})}-1\right).\label{eq:one_particle_distr_z}
\end{align}
We will use this form of one-particle distributions near the interface for our later discussion and refer to it as $g_{sf}$ in the remainder of this work. 

We need to define the fluid pair distribution function near the interface in a different way compared to the one in the bulk. We call $g_{ff}({\bf R}_{fn}^{(0)},\,{\bf R}_{fm}^{(0)})$, the fluid pair distribution function for fluid atoms that are close to the interface. This function depends on both the fluid-fluid and the fluid-solid interaction potential. Following Percus \cite{Percusjk64}, the two-particle probability density $g_{2}({\bf r}_{1},{\bf r}_{2}|U)$ in the presence of an external potential $U$ can be written as 
\begin{gather}
g_{2}({\bf r}_{1},{\bf r}_{2}|U)e^{\beta[U({\bf r}_{1})+U({\bf r}_{2})]}=g_{2}({\bf r}_{1},{\bf r}_{2})+\int {\rm d}^{3}{\bf r}_{3}\mathcal{F}_{3}({\bf r}_{1},{\bf r}_{2},{\bf r}_{3})\left(e^{-\beta U({\bf r}_{3})}-1\right)\nonumber \\
+\frac{1}{2}\int {\rm d}^{3}{\bf r}_{3}\int {\rm d}^{3}{\bf r}_{4}\mathcal{F}_{24}({\bf r}_{1},{\bf r}_{2},{\bf r}_{3},{\bf r}_{4})\left(e^{-\beta U(r_{3})}-1\right)\left(e^{-\beta U(r_{4})}-1\right)+\ldots\label{eq:two_particle_prob_density}
\end{gather}
where $g_{2}({\bf r}_{1},{\bf r}_{2})=g({\bf r}_{1}-{\bf r}_{2})g({\bf r}_{1})g({\bf r}_{2})$. If we ignore three-particle correlations ($\mathcal{F}_{n}=0$ for $n\geq3$), the form of $g_{ff}$ can be approximated as 
\begin{equation}
g_{ff}({\bf R}_{fn}^{(0)},\,{\bf R}_{fm}^{(0)})\approx g(|{\bf R}_{fn}^{(0)}-{\bf R}_{fm}^{(0)}|)e^{-\beta V_{sf}(R_{fn}^{(0)})-\beta V_{sf}(R_{fm}^{(0)})}.\label{eq:gff}
\end{equation} 

\subsubsection{Equations of motion for atoms near the interface}

We assume that the solid atoms are arranged in locations ${\bf R}_{sj}^{(0)}=({\bf \rho}_{j},\, z_{j})$ to the left of the interface (as shown in Fig. \ref{fig:Cartoon_solid_fluid}). The vectors ${\bf \rho}_{j}$ are parallel to the interface, and $\hat{z}$ is perpendicular to the interface. The interface is marked by the plane $z=0$. The solid atoms at the interface have $z_{j}=0$. The solid atoms are displaced by both the incident and the reflected waves. The displacement of an interfacial solid atom (${\bf Q}_{sj}$) can be written as the linear combination:
\begin{equation}
{\bf Q}_{sj}=e^{i({\bf q}\cdot{\bf \rho}_{j}-\omega t)}\left[\hat{\epsilon}_{i}e^{iq_{i}z_{j}}I_{i}+\sum_{\lambda=1}^{3}R_{\lambda}\hat{\epsilon}_{s\lambda}e^{-iq_{s\lambda}z_{j}}\right],\label{eq:solid_displacement}
\end{equation}
where ${\bf q}$ and $\omega$ are the component of the wavevector parallel to the interface and the frequency of both the incident and reflected waves, respectively. The perpendicular component of the wavevector and the polarization vector of the incident and reflected waves are given by ($q_i, \hat{\epsilon}_i$) and ($q_{s\lambda}, \hat{\epsilon}_{s\lambda}$), respectively. In general, the polarization vectors of the three reflected waves, $\hat{\epsilon}_{s\lambda}$ are not mutually orthogonal. Here, $I_{i}$ is the amplitude of the incident wave and $R_{\lambda}$ is the amplitude of the reflected wave for polarization $\lambda$. On the other hand, the displacement of a fluid atom close to the interface (${\bf u}_{fm}$) from its equilibrium location ${\bf R}_{fm}^{(0)}=({\bf \rho}_{fm},\, z_{fm})$ can be written as
\begin{equation}
{\bf u}_{fm}=e^{i({\bf q}\cdot{\bf \rho}_{fm}-\omega t)}\sum_{\lambda=1}^{3}T_{\lambda}\hat{\epsilon}_{f\lambda}e^{-iq_{f\lambda}z_{fm}},\label{eq:fluid_displacement}
\end{equation}
where ${\bf q}$ and $\omega$ are the component of the wavevector parallel to the interface and the frequency of the transmitted waves, respectively. The perpendicular component of the wavevector and the polarization vector of the transmitted waves are given by ($q_{f\lambda}, \hat{\epsilon}_{f\lambda}$), respectively. Here, $T_{\lambda}$ is the amplitude of the transmitted wave for polarization $\lambda$. 
 
The equations of motion for the solid atoms near the interface contain force-terms due to solid-solid interactions as well as force-terms due to solid-fluid interactions \cite{Mahan_kapitza}: 
\begin{align}
\underset{\text{solid atom at \ensuremath{z=0}}}{\underbrace{M_{s}\omega^{2}{\bf Q}_{sj}}}= & \underset{\substack{\text{interaction with}\\ \text{solid atoms with \ensuremath{z<0}}}}{\sum_{\delta_{s}}\underbrace{K(\delta_{s})\hat{\delta}_{s}\hat{\delta}_{s}\cdot({\bf Q}_{sj}-{\bf Q}_{s,j+\delta_{s}})}}\nonumber \\
+ & \underset{\text{interaction with fluid atoms with \ensuremath{z>0}}}{\sum_{\delta}\underbrace{[A_{sf}(\delta)({\bf Q}_{sj}-{\bf u}_{fm})+B_{sf}(\delta)\hat{\delta}\hat{\delta}\cdot({\bf Q}_{sj}-{\bf u}_{fm})]}}\label{eq:solid_atom_interface_equation},
\end{align}
with ${\bf \delta}_{s}={\bf R}_{sj}^{(0)}-{\bf R}_{s,j+\delta_{s}}^{(0)}\text{ and }{\bf \delta}={\bf R}_{sj}^{(0)}-{\bf R}_{fm}^{(0)}$, $M_s$ is mass of the solid atom and $K(\delta_{s})$ is the bond-directed spring constant. The functions $A$ and $B$ are the derivatives of the interaction potential as defined in Eq. (\ref{eq:A}) and (\ref{eq:B}), respectively. The summations over $\delta_{s}$ and $\delta$ include the neighboring solid and fluid atoms near the interface, respectively. We refer to this equation as the solid interface-equation of motion in this article. Similarly, the equations of motion for the fluid atoms near the interface have force-terms due to both fluid-fluid and fluid-solid interactions \cite{Mahan_kapitza}, 
\begin{align}
\underset{\text{fluid atom with }z\approx0}{\underbrace{M_{f}\omega^{2}{\bf u}_{fn}}}= & \underset{\text{interaction with solid atoms with }z<0}{\sum_{\delta}\underbrace{[A_{sf}(\delta)({\bf u}_{fn}-{\bf Q}_{sj})+B_{sf}(\delta)\hat{\delta}\hat{\delta}\cdot({\bf u}_{fn}-{\bf Q}_{sj})]}}\nonumber \\
+ & \underset{\text{interaction with fluid atoms with }z>0}{\sum_{\delta_f}\underbrace{[A_{ff}(\delta_{f})({\bf u}_{fn}-{\bf u}_{fm})+B_{ff}(\delta_{f})\hat{\delta}_{f}\hat{\delta}_{f}\cdot({\bf u}_{fn}-{\bf u}_{fm})]}}\label{eq:liquid_atom_interface_equation},
\end{align}
with ${\bf \delta}={\bf R}_{sj}^{(0)}-{\bf R}_{fn}^{(0)}\text{ and }{\bf \delta}_{f}={\bf R}_{fm}^{(0)}-{\bf R}_{fn}^{(0)}$, $M_f$ is the mass of the fluid atom. The summation over $\delta$ includes the solid atoms close to the fluid atom, near the interface. The summation over $\delta_f$ includes the neighboring fluid atoms near the interface. We refer to this equation as the fluid interface-equation of motion in the article.

\subsubsection{Coupling matrix: $\mathcal{M}$}

In order to evaluate the solid-fluid and the fluid-fluid interaction terms, we need to account for the fact that the positions of the fluid atoms change as they move around.  
To incorporate this, we replace the summations over fluid atoms in Eq.s (\ref{eq:solid_atom_interface_equation})
and (\ref{eq:liquid_atom_interface_equation}) by integrations over all positions of the fluid atoms. Using the one particle distribution function near the interface $g_{sf}$, we reformulate the solid interface-equation of motion (Eq. (\ref{eq:solid_atom_interface_equation})) to obtain \cite{Mahan_kapitza}
\begin{gather}
\underset{\text{solid atoms with \ensuremath{z>0}}}{\sum_{\delta_{s}}\underbrace{K(\delta_{s})\hat{\delta}_{s}\hat{\delta}_{s}\cdot({\bf Q}_{sj}-{\bf Q}_{s,j+\delta_{s}})}}=\mathcal{M}(0,0)\cdot{\bf Q}_{sj}-\mathcal{M}({\bf q},q_{z})\cdot{\bf u}(0),\label{eq:K_M_equation}\\
\text{with \,\,\,}\mathcal{M}(0,0)=n\int_{z>0}{\rm d}^{3}{\bf R}g_{sf}(z)[A_{sf}(R)\mathcal{I}+B_{sf}(R)\hat{R}\hat{R}],\label{eq:M00_integral_form}\\
\text{and\,\,\,}\mathcal{M}({\bf q},q_{z})\cdot{\bf u}(0)=\sum_{\lambda=1}^{3}\mathcal{M}({\bf q},q_{f\lambda})\cdot\hat{\epsilon}_{f\lambda}T_{\lambda}\nonumber \\
\text{with \,\,\,}\mathcal{M}({\bf q},q_{f\lambda})=n\int_{z>0}{\rm d}^{3}{\bf R}g_{sf}(z) [A_{sf}(R)\mathcal{I}+B_{sf}(R)\hat{R}\hat{R}]e^{i{\bf q}\cdot{\bf \rho}+iq_{f\lambda}z},\label{eq:Mkkz_integral_form}
\end{gather}
where $\mathcal{I}$ is the unit tensor and $({\bf q},q_{f\lambda})$ are the wave vectors of the three transmitted waves in fluid. Both the solid-fluid coupling matrices $\mathcal{M}(0,0)$ and $\mathcal{M}({\bf q},q_{z})$ are second rank tensors, one can write them in component form as shown in Eq. (44)-(54) from Reference \citenum{Mahan_kapitza}.

\subsubsection{Coupling matrix: $\mathcal{U},\,\mathcal{V}$}

We reformulate the fluid interface-equation of motion (Eq.(\ref{eq:liquid_atom_interface_equation})) in a similar way. Introducing the pair distribution functions, we
obtain
\begin{gather}
\underset{\text{fluid atom with }z\approx0}{\underbrace{M_{f}\omega^{2}{\bf u}_{fn}}}=\underset{\text{interaction with solid atoms with }z<0}{\sum_{j}\underbrace{[A_{sf}(\delta)({\bf u}_{fn}-{\bf Q}_{sj})+B_{sf}(\delta)\hat{\delta}\hat{\delta}\cdot({\bf u}_{fn}-{\bf Q}_{sj})]}}\nonumber \\
+\underset{\text{interaction with fluid atoms with }z>0}{\underbrace{n\int {\rm d}^{3}{\bf R}g_{ff}({\bf R}_{fn}^{(0)},{\bf R})[A_{ff}(\delta_{f})({\bf u}_{fn}-{\bf u}({\bf R}))+B_{ff}(\delta_{f})\hat{\delta}_{f}\hat{\delta}_{f}\cdot({\bf u}_{fn}-{\bf u}({\bf R}))]}}\label{eq:liquid_atom_interface_equation_pdf}\\
\text{with }{\bf \delta}={\bf R}_{sj}^{(0)}-{\bf R}_{fn}^{(0)}\text{ and }{\bf \delta}_{f}={\bf R}-{\bf R}_{fn}^{(0)}.\nonumber 
\end{gather}
We replace the LHS with the bulk-equation of motion of the fluid atoms Eq. (\ref{eq:fluid_equation_of_motion}) and obtain 
\begin{gather}
\underset{\text{interaction with fluid atoms with }z<0}{\underbrace{n\int {\rm d}^{3}{\bf R}g(|{\bf R}_{fn}^{(0)}-{\bf R}|)[A_{ff}(\delta_{f})\mathcal{I}+B_{ff}(\delta_{f})\hat{\delta}_{f}\hat{\delta}_{f}]\cdot({\bf u}_{fn}-{\bf u}({\bf R}))}}\nonumber \\
+\underset{\text{interaction with fluid atoms with }z>0}{\underbrace{n\int {\rm d}^{3}{\bf R}(g(|{\bf R}_{fn}^{(0)}-{\bf R}|)-g_{ff}({\bf R}_{fn}^{(0)},{\bf R}))[A_{ff}(\delta_{f})\mathcal{I}+B_{ff}(\delta_{f})\hat{\delta}_{f}\hat{\delta}_{f}]\cdot({\bf u}_{fn}-{\bf u}({\bf R}))}}\nonumber \\
=\underset{\text{interaction with solid atoms with }z<0}{\sum_{j}\underbrace{[A_{sf}(\delta)+B_{sf}(\delta)\hat{\delta}\hat{\delta}]\cdot({\bf u}_{fn}-{\bf Q}_{sj})}},\label{eq:fluid_bulk_minus_interface_gff}
\end{gather}
where $g(\left|{\bf R}_{fn}^{(0)}-{\bf R}\right|)$ and $g_{ff}({\bf R}_{fn}^{(0)},{\bf R})$ are the fluid pair distribution functions in the bulk and near the interface, respectively. $g_{ff}({\bf R}_{fn}^{(0)},\,{\bf R})$ is a function of the position vectors of both fluid atoms, as a result it is difficult to compute the integral with $g_{ff}$ in the integrand. We make the simplifying assumption that the pair distribution function of the fluid atoms near the interface do not differ much from the bulk pair distribution function,
so that $g(|{\bf R}_{fn}^{(0)}-{\bf R}|)\approx g_{ff}({\bf R}_{fn}^{(0)},{\bf R})$ in the first approximation and the second term in the LHS vanishes in the above equation (Eq. (\ref{eq:fluid_bulk_minus_interface_gff})). Hence, we have
\begin{gather}
\underset{\text{interaction with fluid atoms with }z<0}{\underbrace{n\int {\rm d}^{3}{\bf R}g(|{\bf R}_{fn}^{(0)}-{\bf R}|)[A_{ff}(\delta_{f})\mathcal{I}+B_{ff}(\delta_{f})\hat{\delta}_{f}\hat{\delta}_{f}]\cdot({\bf u}_{fn}-{\bf u}({\bf R}))}}\nonumber \\
=\underset{\text{interaction with solid atoms with }z<0}{\sum_{j}\underbrace{[A_{sf}(\delta)+B_{sf}(\delta)\hat{\delta}\hat{\delta}]\cdot({\bf u}_{fn}-{\bf Q}_{sj})}}.\label{eq:fluid_bulk_minus_interface}
\end{gather} 
In order to incorporate the effect of the changing position of the fluid atom, we multiply both sides of the equation with $n\times g_{sf}(z_{fn})$ and average over all possible positions. Integrating with respect to ${\bf R}_{fn}$ we obtain 
\begin{align}
&n^{2}\int_{z>0}{\rm d}^{3}{\bf R}_{fn}\ g_{sf}(z_{fn})\int_{z<0}{\rm d}^{3}{\bf R}\ g(|{\bf R}_{fn}-{\bf R}|)\nonumber \\
&\times[A_{ff}(|{\bf R}_{fn}-{\bf R}|)\mathcal{I}+B_{ff}(|{\bf R}_{fn}-{\bf R}|)\frac{({\bf R}-{\bf R}_{fn})^{2}}{|{\bf R}-{\bf R}_{fn}|^{2}}]\cdot({\bf u}_{fn}-{\bf u}({\bf R}))\nonumber \\
&=\sum_{j,z<0}n\int_{z>0}{\rm d}^{3}{\bf R}_{fn}\ g_{sf}(z_{fn})\label{eq:fluid_bulk_minus_interface_gusf}\\
&\times[A_{sf}(|{\bf R}_{fn}-{\bf R}_{sj}|)+B_{sf}(|{\bf R}_{fn}-{\bf R}_{sj}|)\frac{({\bf R}_{fn}-{\bf R}_{sj})^{2}}{|{\bf R}_{fn}-{\bf R}_{sj}|^{2}}]\cdot({\bf u}_{fn}-{\bf Q}_{sj}).\nonumber 
\end{align}
Renaming ${\bf R}_{fn}={\bf R}^{\prime}$ and changing the variables
from (${\bf R}^{\prime}$, ${\bf R}$) to (${\bf R}^{\prime}$, ${\bf R}^{\prime}-{\bf R}={\bf r}$)
in the LHS of the above equation and from ${\bf R}^{\prime}$ to (${\bf R}^{\prime}-{\bf R}_{sj}={\bf r}$)
in the RHS, we obtain using ${\bf q}_{T}=({\bf q},q_{f\lambda})$
\begin{align}
&\sum_{\lambda=1}^{3} T_{\lambda}\hat{e}_{f\lambda}\cdot\left(n^{2}\int_{z>0}{\rm d}^{3}R^{\prime}g_{sf}(Z^{\prime})\int_{\text{all space}}{\rm d}^{3}rg(r)\right.\nonumber \\
& \times \left.[A_{ff}(r)\mathcal{I}+B_{ff}(r)\hat{r}\hat{r}](e^{i{\bf q}_{T}.{\bf R}^{\prime}}-e^{i{\bf q}_{T}.({\bf R}^{\prime}-{\bf r})})\right)\nonumber \\
&=\sum_{j,z<0}n\int_{z>0}{\rm d}^{3}rg_{sf}(|{\bf r}+{\bf R}_{sj}|_z)[A_{sf}(r)\mathcal{I}+B_{sf}(r)\hat{r}\hat{r}]\ \cdot \nonumber \\
&\left(\sum_{\lambda=1}^{3} T_{\lambda}\hat{e}_{f\lambda}e^{i{\bf q}_{T}.({\bf r}+{\bf R}_{sj})}-e^{i{\bf q}.{\bf \rho}_{sj}}\left[\hat{e}_{i}e^{iq_{i}z_{sj}}I_{i}+\sum_{\lambda=1}^{3}R_{\lambda}\hat{e}_{s\lambda}e^{-iq_{s\lambda}z_{sj}}\right]\right).\label{eq:U_V_1}
\end{align}
The fluid atoms that contribute most to the solid-fluid interaction term in the above equation are the ones that are closest to the solid atom at the origin. We assume that ${\bf R}_{sj}\approx(0,0,0)$ and obtain 
\begin{align}
&\sum_{\lambda=1}^{3} T_{\lambda}\hat{e}_{f\lambda}\cdot\left(n^{2}\int_{z>0}{\rm d}^{3}R^{\prime}g_{sf}(Z^{\prime})\int_{\text{all space}}{\rm d}^{3}rg(r)\right.\nonumber \\
&\times \left.[A_{ff}(r)\mathcal{I}+B_{ff}(r)\hat{r}\hat{r}](e^{i{\bf q}_{T}.{\bf R}^{\prime}}-e^{i{\bf q}_{T}.({\bf R}^{\prime}-{\bf r})})\right)\nonumber \\
&=n\int_{z>0}{\rm d}^{3}rg_{sf}(z)[A_{sf}(r)\mathcal{I}+B_{sf}(r)\hat{r}\hat{r}]\ \cdot\nonumber \\
&\left(\sum_{\lambda=1}^{3} T_{\lambda}\hat{e}_{f\lambda}e^{i{\bf q}_{T}.{\bf r}}-\left[\hat{e}_{i}I_{i}+\sum_{\lambda=1}^{3}R_{\lambda}\hat{e}_{s\lambda}\right]\right).\label{eq:U_V_2}
\end{align}
We can write the interface equation in a compact form as
\begin{equation}
[\mathcal{U}({\bf q},q_{z})-\mathcal{V}({\bf q},q_{z})]\cdot{\bf u}(0)=\mathcal{M}({\bf q},q_{z})\cdot{\bf u}(0)-\mathcal{M}(0,0)\cdot{\bf Q}_{sj},\label{eq:U_V_equation}
\end{equation}
where $\mathcal{U}$ and $\mathcal{V}$ are the fluid-solid coupling
matrices
\begin{eqnarray}
\mathcal{U}({\bf q},q_{fn}) & = & n^{2}\int_{z>0}{\rm d}^{3}R^{\prime}g_{sf}(Z^{\prime})e^{i{\bf q}_{T}.{\bf R}^{\prime}}\int_{\text{all space}}{\rm d}^{3}rg(r)[A_{ff}(r)+B_{ff}(r)\hat{r}\hat{r}],\label{eq:U_integral_form}\\
\mathcal{V}({\bf q},q_{fn}) & = & n^{2}\int_{z>0}{\rm d}^{3}R^{\prime}g_{sf}(Z^{\prime})e^{i{\bf q}_{T}.{\bf R}^{\prime}}\int_{\text{all space}}{\rm d}^{3}rg(r) e^{-i{\bf q}_{T}.{\bf r}}[A_{ff}(r)+B_{ff}(r)\hat{r}\hat{r}],\label{eq:V_integral_form}
\end{eqnarray}
and $\mathcal{M}$ was defined in Eq. (\ref{eq:Mkkz_integral_form}). Both $\mathcal{U}$ and $\mathcal{V}$ are second-rank tensors and we can write them in component form as follows:
\begin{eqnarray}
\mathcal{U}({\bf q},q_{z}) & = & U_{A}({\bf q},q_{f\lambda})\mathcal{I}+U_{\perp}({\bf q},q_{f\lambda})(\hat{x}\hat{x}+\hat{y}\hat{y})+\hat{z}\hat{z}U_{z}({\bf q},q_{f\lambda}),\label{eq:U_component_form}\\
\mathcal{V}({\bf q},q_{z}) & = & V_{A}({\bf q},q_{f\lambda})\mathcal{I}+\hat{x}\hat{x}V_{xx}({\bf q},q_{f\lambda})+\hat{y}\hat{y}V_{yy}({\bf q},q_{f\lambda})\nonumber \\
 & + & \hat{z}\hat{z}V_{zz}({\bf q},q_{f\lambda})+(\hat{x}\hat{y}+\hat{y}\hat{x})V_{xy}({\bf q},q_{f\lambda})\nonumber \\
 & + & (\hat{x}\hat{z}+\hat{z}\hat{x})V_{xz}({\bf q},q_{f\lambda})+(\hat{z}\hat{y}+\hat{y}\hat{z})V_{zy}({\bf q},q_{f\lambda}),\label{eq:V_component_form}
\end{eqnarray}
where the different components of $\mathcal{U}({\bf q},q_{f\lambda})$ in cylindrical polar coordinates are given by:
\begin{gather}
U_{A}=I_{surface}\times\left(2\pi n\int_{-\infty}^{\infty}{\rm d}z\int_{0}^{\infty} {\rm d}\rho \rho g(r)A_{ff}(r)\right),\nonumber \\
U_{\perp}=I_{surface}\times\left(\pi n\int_{-\infty}^{\infty}{\rm d}z\int_{0}^{\infty} {\rm d}\rho \rho g(r)B_{ff}(r)\frac{\rho^{2}}{\rho^{2}+z^{2}}\right),\label{eq:UA_Uperp_Uz}\\
U_{z}=I_{surface}\times\left(2\pi n\int_{-\infty}^{\infty}{\rm d}z\int_{0}^{\infty} {\rm d}\rho \rho g(r)B_{ff}(r)\frac{z^{2}}{\rho^{2}+z^{2}}\right),\nonumber \\
\text{with }I_{surface}=2\pi n\int_{0}^{\infty}{\rm d}z^{\prime}\exp[iq_{f\lambda}z^{\prime}]g_{sf}(z^{\prime})\int_{0}^{\infty}{\rm d}\rho^{\prime}\rho^{\prime}J_{0}(q\rho^{\prime}),\label{eq:int_surface}
\end{gather}
where $J_{0}(q\rho)$ is the Bessel function of the first kind. The set of tensor elements ($V_{xx}$, $V_{yy}$, $V_{zz}$, $V_{xy}$, $V_{yz}$, $V_{xz}$) have each terms with $A_{ff}$ and $B_{sff}$ in the integrand. Grouping together similar terms, we can write 
\begin{eqnarray}
\mathcal{V}({\bf q},q_{f\lambda}) & = & V_{A}({\bf q},q_{f\lambda})\mathcal{I}+(\hat{x}\hat{x}+\hat{y}\hat{y})V_{0\perp}({\bf q},q_{f\lambda})+\hat{z}\hat{z}V_{0z}({\bf q},q_{f\lambda})\nonumber \\
 & - & [(\hat{x}\hat{x}-\hat{y}\hat{y})\cos(2\phi_{q})+(\hat{x}\hat{y}+\hat{y}\hat{x})\sin(2\phi_{q})]V_{2\perp}({\bf q},q_{f\lambda})\label{eq:Vkkz_grouped_component_form}\\
 & + & i[(\hat{x}\hat{z}+\hat{z}\hat{x})\cos(\phi_{q})+(\hat{z}\hat{y}+\hat{y}\hat{z})i\sin(\phi_{q})]V_{1}({\bf q},q_{f\lambda}).\nonumber 
\end{eqnarray}
where $\phi_{q}$ is the angle between the component of the wavevector parallel to the interface and the $x$-axis, ${\bf q}=q[\cos(\phi_{q}),\sin(\phi_{q})]$. The different elements of the tensor are given by 
\begin{gather}
V_{A}({\bf q},q_{f\lambda})=I_{surface}\times\left(2\pi n\int_{-\infty}^{\infty}{\rm d}z\exp[-iq_{f\lambda}z]\int_{0}^{\infty} {\rm d}\rho \rho g(r)A_{ff}(r)J_{0}(q\rho)\right),\nonumber \\
V_{l\perp(l=0,2)}({\bf q},q_{f\lambda})=I_{surface}\times\left(\pi n\int_{-\infty}^{\infty}{\rm d}z\exp[-iq_{f\lambda}z]\int_{0}^{\infty}{\rm d}\rho \rho^{3} g(r)\frac{B_{ff}(r)}{r^{2}}J_{l}(q\rho)\right),\label{eq:VA_Vlperp_V0z_V1}\\
V_{0z}({\bf q},q_{f\lambda})=I_{surface}\times\left(2\pi n\int_{-\infty}^{\infty}z^{2}{\rm d}z\exp[-iq_{f\lambda}z]\int_{0}^{\infty} {\rm d}\rho \rho g(r)\frac{B_{ff}(r)}{r^{2}}J_{0}(q\rho)\right),\nonumber \\
V_{1}({\bf q},q_{f\lambda})=I_{surface}\times\left(-2\pi n\int_{-\infty}^{\infty}z{\rm d}z\exp[-iq_{f\lambda}z]\int_{0}^{\infty}{\rm d}\rho \rho^{2}g(r)\frac{B_{ff}(r)}{r^{2}}J_{1}(q\rho)\right).\nonumber 
\end{gather}
The tensors $\mathcal{U}$ and $\mathcal{V}$ determine the coupling between the atoms in the fluid. Both of them have spring constant units. Using the form of the LJ potential (Eq. (\ref{eq:LJ_potential})) parts of the integrals in the different elements of the tensors $\mathcal{M}$, $\mathcal{U}$ and $\mathcal{V}$ can be evaluated analytically. The rest of the integration needs to be performed numerically. 

\section{Kapitza conductance}
\label{KC}

When a phonon is incident to the solid-fluid interface, there can be up to three reflected and three transmitted waves to which the incident phonon can transfer energy. The three reflection and three transmission amplitudes ($\frac{R_{\lambda}}{I_i},\frac{T_{\lambda}}{I_i}$) are introduced in Eq. (\ref{eq:solid_displacement}) and Eq. (\ref{eq:fluid_displacement}), respectively. Equations (\ref{eq:K_M_equation}) and (\ref{eq:U_V_equation}) are the two key equations, which describe the motion of interfacial solid and fluid atoms. These two equations include the six amplitudes as unknowns. Each of these two equations is a vector equation with three components, and the six equations form a coupled $6\times6$ set of linear equations. The six equations can be expressed in matrix form as
\begin{equation}
\sum_{\lambda=1}^{3}C_{p\lambda}R_{\lambda}+\sum_{\lambda=4}^{6}C_{p\lambda}T_{\lambda-3}=C_{p0}I_{i},\ p=1,\ldots6.\label{eq:Matrix_equation}
\end{equation}
The coefficients $C_{p\lambda}$ are given in the Appendix. Solving these coupled linear equations, we obtain the reflection and transmission amplitudes. The transmission amplitudes are used to calculate the phonon transmission coefficients. For an incident phonon of frequency $\omega_{\lambda}({\bf q})$, wavevector ${\bf q}$, and polarization $\lambda$ the coefficient of transmission into all possible modes, $\mathcal{T}_{\lambda}({\bf q})$, is given by
\begin{equation}
\mathcal{T}_{\lambda}({\bf q})=\frac{\rho_f\sum_{\lambda^{\prime}=1}^3 v^{(f)}_{\lambda^{\prime}z}|T_{\lambda^{\prime}}|^{2}}{\rho_s v^{(s)}_{\lambda z}|I_{\lambda}|^{2}}.\label{eq:transmission_coeff}
\end{equation}
Here, $\rho_s$ and $\rho_f$ are the mass densities of the solid and the fluid, respectively. $v^{(s)}_{\lambda z}$ is the $z$ component of the group velocity of the incident phonon and $v^{(f)}_{\lambda^{\prime} z} (\lambda^{\prime}=1,2,3)$ are the group velocities of the transmitted sound waves in the fluid.  

\section{Illustrative example}
\label{example}

We chose the FCC solid argon-fluid neon interface as an example to illustrate the calculation of the Kapitza conductance using our lattice dynamical model. This particular example was chosen because this system is representative of a solid-fluid interface where the two systems are interacting via a central potential. We define the dimensionless length variables of the solid-fluid system in the following way: 
\begin{eqnarray}
r^{*}=\frac{r}{\sigma_{sf}}&, & q^{*}=q\sigma_{sf}; \nonumber \\
n^{*}=n\sigma_{sf}^{3} &,& T^{*}=\frac{k_{B}T}{\epsilon_{sf}}.\label{eq:dimensionless_variables}
\end{eqnarray}
The values of the LJ potential parameters for the solid-fluid system are shown in Table \ref{tab:LJ_parameters}. 
\begin{table}
\begin{centering}
\begin{tabular}{|c|c|c|}
\hline 
 & $\sigma$($\text{\AA}$) & $\epsilon$(K)\tabularnewline
\hline 
\hline 
Ne-Ne \cite{Kittel} & 2.740 & 36.23\tabularnewline
\hline 
Ne-Ar \cite{Klein} & 3.083 & 64.50\tabularnewline
\hline 
\end{tabular}
\par\end{centering}
\caption{Lennard-Jones potential parameters for Ne-Ne and Ne-Ar interaction.
\label{tab:LJ_parameters}}
\end{table}

Under normal pressure, neon remains liquid in a very narrow temperature range, $24.56$K - $27.07$K. This puts some constraint over the choice of thermodynamic parameters for the solid-fluid system. 
We have found that the parameters $n$ = 0.0239 $\text{\AA}^{-3}$ and $T$ = 54.345 K ensure fast convergence of the integral equations and thereby evaluate the pair distribution function. Argon remains solid at this temperature.

\begin{figure}
\begin{centering}
\includegraphics[scale=0.4]{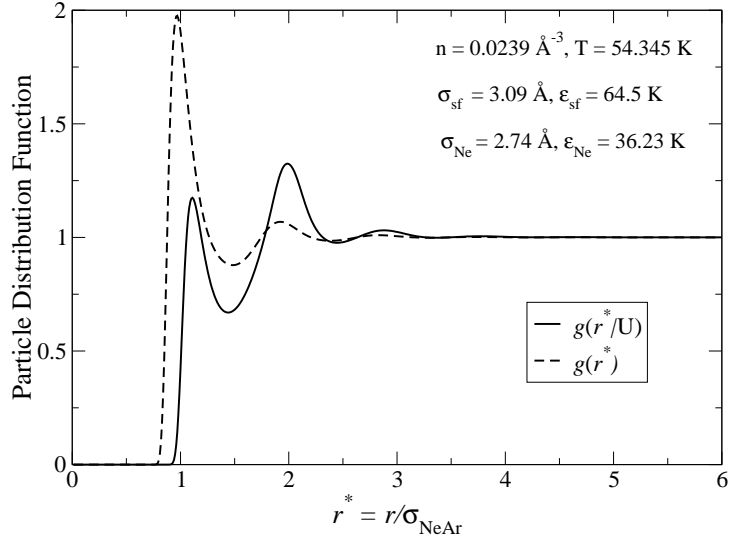}
\par\end{centering}
\caption{Particle distribution functions of fluid neon obtained using integral equation theories. The dashed curve represents the pair distribution function of neon atoms in the bulk fluid, evaluated using the hypernetted chain approximation, Eq. (\ref{eq:HNC_equation}) \cite{HNC}. The solid curve represents the one particle distribution function of neon atoms near the solid argon interface, evaluated using Eq. (\ref{eq:g_sf}). The length variable is scaled with respect to the LJ argon-neon length parameter $\sigma_{\text{Ne-Ar}}$. \label{fig:pdfs}}
\end{figure}

\begin{figure}
\begin{centering}
\includegraphics[scale=0.4]{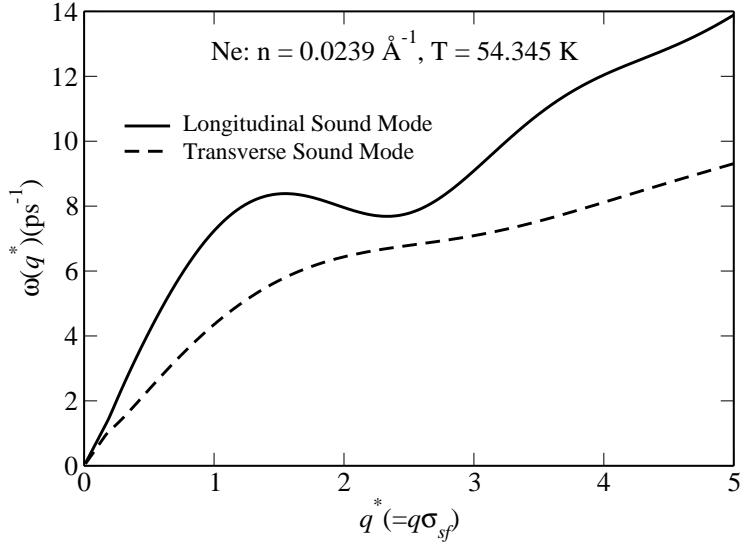}
\par\end{centering}
\caption{Dispersion of longitudinal and transverse sound waves in liquid neon
at $n=0.0239$\AA$^{-3}$ and $T=54.345$K. The length variable is scaled with respect to the LJ argon-neon length parameter $\sigma_{\text{Ne-Ar}}$. \label{fig:Ne_sound_modes} }
\end{figure}

The phonon dispersion in solid argon is shown in Fig. \ref{fig:solid_phonons}. The bulk pair distribution function of fluid neon for the particular choice of thermodynamic parameters is shown by the dashed line in Fig. \ref{fig:pdfs}. 
We then numerically integrate the dispersion relations (Eq. (\ref{eq:longitudinal_freq})) to obtain the dispersion of sound waves in bulk fluid neon. The dispersion of sound waves in fluid neon are shown in Fig. \ref{fig:Ne_sound_modes} for a range of $q$ values. 

As a consequence of the solid-fluid interaction, the distribution of neon atoms near the solid argon interface is different from that in the bulk. The one-particle distribution function of neon atoms near the solid argon interface, evaluated using Eq. (\ref{eq:one_particle_distr}), is shown by the solid line in Fig. \ref{fig:pdfs}. We select 1000 randomly generated points in the first brillouin zone of solid argon as incident phonons. 
We disregard points close to one of the symmetry points of the FCC lattice. The reflected and transmitted waves for an incident phonon are determined such that the frequency and parallel components of the wavevector of all the modes are identical. The detailed procedure is given in Subsection \ref{mode_matching}. We solve the coupled equations of motion of the interfacial solid and the fluid atoms for the choice of incident phonons to obtain the phonon reflection and transmission amplitudes. 
The values of the transmission coefficients when inserted into Eq. (\ref{eq:final_KR}), gives an estimate of the interfacial thermal conductance. Inserting these values in Eq. (\ref{eq:final_KR}), we obtain the value of the Kapitza conductance of solid argon-fluid neon interface, $G=$ 37.4 MW$\text{K}^{-1}\text{m}^{-2}$. Although we could not compare our result with any experimental results reported for a similar system, we note that our estimate lies within the range of experimental values of the Kapitza conductance measured for solid metal-liquid water interfaces \cite{Ge}.

\section{Summary and discussion}
\label{conclusion}

We propose a novel lattice dynamical model calculation to estimate the Kapitza conductance of insulating solid-fluid interfaces incorporating full interfacial structure details. The existing theoretical models, namely AMM and DMM only consider bulk system properties to estimate the Kapitza conductance. We believe our model is an improvement over these two widely used models, due to the fact that it includes detailed properties of the interface to compute the Kapitza conductance of solid-fluid interfaces. In our model, we consider that heat is mostly carried by phonons in the solid side, and that the sound waves with matched frequencies carry diffusive heat into the bulk of the liquid from the interface. 
The dispersion of the sound waves as well as the interfacial fluid structure are determined using approximate integral equation theories (Percus-Yevick, HNC). 
The coupled equations of motion of the interfacial solid and fluid atoms yield the phonon reflection and the transmission coefficients at the interface. The Kapitza conductance of the solid-fluid interface is obtained from the phonon transmission coefficients at the interface. 
The only input parameters required in our model are the pairwise interaction potential parameters. Once the bulk system properties are determined and the interfacial potential parameters are known, the rest of the calculation does not depend on any further input. To ensure convergence of the integral equation theories within reasonable computational time, one needs to be careful when choosing the thermodynamic parameters of the fluid. However, this restriction can be removed if one uses other methods (e.g. molecular dynamics) to 
compute the fluid pair distribution function and uses the data as input in the model. We made some approximations while calculating the interfacial fluid distribution functions to ease the computation. However, one could incorporate a more detailed interfacial fluid distribution to improve the estimate.  As an illustrative example, we derived the Kapitza conductance of solid argon-fluid neon interface, using pairwise Lennard-Jones interaction potentials. Our method provides a reasonable estimate of Kapitza conductances at insulating solid-fluid interfaces. Though no experimental data is readily available, our results should provide a solid estimate for future studies. 

Our model can be generalised to obtain the Kapitza conductance of generic hard matter-soft matter interfaces (e.g. solid-polymer interface, solid-amorphous material interface). Hence, this method provides opportunities to obtain important insight into thermal transport in heterogeneous nanostructured systems incorporating generic hard matter-soft matter interfaces (e.g. solid-polymer interface, solid-amorphous material interface). A detailed understanding of heat transport in heterogeneous materials is highly relevant for many technological applications including thermal management at the nanoscale and efficient conversion of waste heat into electricity. Therefore, we hope that our investigation will not only help to further advance the understanding of the vibrational energy transport in multicomponent systems, but will guide the modeling and the design of hybrid systems with tailored thermal transport properties for a wide range of technological applications as well.

\acknowledgements
We thank Ganesh Jaya Sreejith for highly valuable discussions, and Juzar Thingna and Tristan Bereau for critical reading of the manuscript.

\bibliographystyle{prsty}
\bibliography{KR-Biblio-Database}

\section{Appendix}

\subsection{Bulk equations of motion}

\subsubsection{Solid}
In the harmonic approximation, the potential energy between the atoms in the solid can be approximated as 
\begin{gather} 
V_{ss}=\frac{1}{2}\sum_{j,\delta_{s}}K(\delta_{s})[\hat{\delta}_{s}\cdot({\bf Q}_{sj}-{\bf Q}_{s,j+\delta_{s}})]^{2},\label{eq:solid_potential_energy}
\end{gather} 
where ${\bf Q}_{sj}$ represents the displacement of the $j^{\text{th}}$ solid atom from its equilibrium position ${\bf R}_{sj}^{(0)}$ $\left({\bf Q}_{sj}\ll\mbox{{\bf R}}_{sj}^{(0)}\right)$, ${\bf \delta}_{s}$ is the distance between nearest neighbor atoms $\left( = {\bf R}_{sj}^{\text{(0)}}-{\bf R}_{s,j+\delta_{s}}^{\text{(0)}}\right)$ and the spring constant $K$ is bond directed. The equation of motion of a bulk solid atom far from the interface, is given by
\begin{equation}
M_{s}\omega^{2}{\bf Q}_{j}=\sum_{\delta_{s}}K(\delta_{s})\hat{\delta}_{s}\hat{\delta}_{s}\cdot({\bf Q}_{sj}-{\bf Q}_{s,j+\delta_{s}}),\label{eq:solid_equation_of_motion}
\end{equation}
where $M_{s}$ is the mass of a solid atom and $\omega$ is the frequency of the phonon in the lattice. The solution for the equation of motion of the solid atom in the bulk Eq. (\ref{eq:solid_equation_of_motion}) is assumed to be of the form
\begin{equation}
{\bf Q}_{sj}=\hat{e}\exp i\left({\bf q}.{\bf R}-\omega t\right),\label{eq:solid_displacement_solution}
\end{equation}
where $\hat{e}$ is the polarization vector, ${\bf q}$ is the wave vector and $\omega$ is the frequency of the phonon. Substitution of the ansatz Eq. (\ref{eq:solid_displacement_solution}) into the equation of motion Eq. (\ref{eq:solid_equation_of_motion}) yields,
\begin{equation}
\underbar{D}\hat{e}=\omega^{2}\hat{e}.\label{eq:solid_eigenvalue_equation}
\end{equation}
The eigenvalue equation Eq. (\ref{eq:solid_eigenvalue_equation}) has a solution if the secular determinant of the dynamical matrix vanishes,
\begin{equation}
\left|\underbar{D}-\omega^{2}\underbar{I}\right|=0.\label{eq:secular_determinant}
\end{equation}

For this work, we consider a face centered cubic (FCC) lattice, the interface is marked by the $(001)$ plane of atoms. We consider a unit cell with one atom to describe the FCC lattice. The atoms of mass $M_{s}$ in the lattice are connected to their $12$ nearest neighbors by springs of stiffness $K_{1}$ and to their $6$ next-nearest neighbors by springs of stiffness $K_{2}$. The spacing between nearest neighbors is $a\slash\sqrt{2}$. The directional vectors connecting these neighbors are given by ${\bf \delta}_{1}$ $=$ $[\frac{a}{2}(\pm1,\pm1,0)$, $\frac{a}{2}(\pm1,0,\pm1)$, $\frac{a}{2}(0,\pm1,\pm1)]$ and ${\bf \delta}_{2}$ $=$ $[a(\pm1,0,0)$,
$a(0,\pm1,0)$, $a(0,0,\pm1)]$, respectively. The dynamical matrix, $\underbar{D}$, for the FCC lattice can be written as
\begin{equation}
\underbar{D}=\frac{K_{1}}{2M_{s}}\left(\begin{array}{ccc}
D_{11} & D_{12} & D_{13}\\
D_{21} & D_{22} & D_{23}\\
D_{31} & D_{32} & D_{33}
\end{array}\right) \label{eq:solid_dynamical_matrix}
\end{equation}
with
\allowdisplaybreaks 
\begin{eqnarray}
D_{11} & = & 2X-2X\cos(\theta_{x}) \nonumber \\
& + & 4-2\cos\left(\frac{\theta_{x}}{2}\right)\cos\left(\frac{\theta_{y}}{2}\right)-2\cos\left(\frac{\theta_{x}}{2}\right)\cos\left(\frac{\theta_{z}}{2}\right),\nonumber \\
D_{12} & = & D_{21}=2\sin\left(\frac{\theta_{x}}{2}\right)\sin\left(\frac{\theta_{y}}{2}\right),\nonumber \\
D_{13} & = & D_{31}=2\sin\left(\frac{\theta_{x}}{2}\right)\sin\left(\frac{\theta_{z}}{2}\right), \nonumber \\
D_{22} & = & 2X-2X\cos(\theta_{y}) \\
& + & 4-2\cos\left(\frac{\theta_{x}}{2}\right)\cos\left(\frac{\theta_{y}}{2}\right)-2\cos\left(\frac{\theta_{y}}{2}\right)\cos\left(\frac{\theta_{z}}{2}\right),\nonumber \\
D_{23} & = & D_{32}=2\sin\left(\frac{\theta_{y}}{2}\right)\sin\left(\frac{\theta_{z}}{2}\right),\nonumber \\
D_{33} & = & 2X-2X\cos(\theta_{z}) \nonumber \\
& + & 4-2\cos\left(\frac{\theta_{x}}{2}\right)\cos\left(\frac{\theta_{z}}{2}\right)-2\cos\left(\frac{\theta_{y}}{2}\right)\cos\left(\frac{\theta_{z}}{2}\right),\nonumber
\end{eqnarray}
where $\theta_{i}\equiv q_{i}a$, $X\equiv\frac{K_{2}}{K_{1}}$. For given values of the wavevector ${\bf q}$, we solve the eigenvalue equation Eq. (\ref{eq:secular_determinant}) to obtain the phonon dispersion in the FCC solid.

\subsubsection{Fluid}

We assume that the fluid atoms are spherically symmetric and interact with each other via a central potential, $V_{ff}(r)$. We also assume that the fluid atoms undergo small displacements, ${\bf u}_{f}$, from their equilibrium positions, ${\bf R}_{f}^{(0)}$, during the propagation of the sound waves. We can then approximate the interaction potential between the fluid atoms as:
\begin{gather}
V_{ff}(|{\bf R}_{fn}-{\bf R}_{fm}|)=V(|{\bf R}_{fn}^{(0)}-{\bf R}_{fm}^{(0)}|)+({\bf u}_{fn}-{\bf u}_{fm})\cdot{\bf F}_{ff}(|{\bf R}_{fn}^{(0)}-{\bf R}_{fm}^{(0)}|)\label{eq:fluid_potential_Taylor_expansion}\\
+\frac{1}{2}\{A_{ff}(R)({\bf u}_{fn}-{\bf u}_{fm})^{2}+B_{ff}(R)[({\bf u}_{fn}-{\bf u}_{fm})\cdot\hat{\delta}_{f}]^{2}\},\nonumber \\
\text{where \,\,\,}A_{ff}(R)=\frac{1}{R}\frac{dV_{ff}}{dR},\ {\bf F}_{ff}(R)={\bf \delta}_{f}A_{ff}\label{eq:Aff}\\
\text{and}\,\,\, B_{ff}(R)=\frac{d^{2}V_{ff}}{dR^{2}}-A_{ff}(R),\label{eq:Bff}\\
\text{with\,\,\,}{\bf \delta}_{f}={\bf R}_{fn}^{(0)}-{\bf R}_{fm}^{(0)},{\bf \hat{\delta}}_{f}=\frac{{\bf \delta}_{f}}{|{\bf \delta}_{f}|}.\label{eq:delta_fluid}
\end{gather}
Here, ${\bf R}_{fn}^{(0)}$ and ${\bf R}_{fm}^{(0)}$ denote the equilibrium positions of the $n^{\text{th}}$ and $m^{\text{th}}$ fluid atoms and ${\bf u}_{fn}$ and ${\bf u}_{fm}$ denote the displacements from their equilibrium positions, respectively. The first-order force term vanishes when we take the average over all fluid atoms. The equation of motion of a bulk fluid atom can then be written as
\begin{align}
M_{f}\omega^{2}{\bf u}_{fn}= & \sum_{m}[A_{ff}(\delta_{f})({\bf u}_{fn}-{\bf u}_{fm})+B_{ff}(\delta_{f})\hat{\delta}_{f}\hat{\delta}_{f}\cdot({\bf u}_{fn}-{\bf u}_{fm})].\label{eq:fluid_equation_of_motion}
\end{align}
Here $M_f$ is the mass of the fluid atom and $\omega$ is the frequency of the sound wave in the fluid. Detailed discussion about this assumption is given in Reference \citenum{Mahan_kapitza}. 

\subsection{Coefficients for coupled solid-fluid interface equations of motion}

The coefficients $C_{p\lambda}$ for the three reflected waves ($\lambda=1,2,3$)
are
\begin{align}
C_{1\lambda} & =K\left[\left(1-(\cos(q_{x}a)\exp(-iq_{z}^{(\lambda)}a))\right)e_{x}^{(\lambda)}-i\sin(q_{x}a)\exp(-iq_{z}^{(\lambda)}a)e_{z}^{(\lambda)}\right]\nonumber \\
 & -\mathcal{M}(0,0)\left[1,1\right]e_{x}^{(\lambda)},\nonumber \\
C_{2\lambda} & =K\left[\left(1-(\cos(q_{y}a)\exp(-iq_{z}^{(\lambda)}a))\right)e_{y}^{(\lambda)}-i\sin(q_{y}a)\exp(-iq_{z}^{(\lambda)}a)e_{z}^{(\lambda)}\right]\nonumber \\
 & -\mathcal{M}(0,0)\left[2,2\right]e_{y}^{(\lambda)},\nonumber \\
C_{3\lambda} & =K\left[-i\sin(q_{x}a)\exp(-iq_{z}^{(\lambda)}a)e_{x}^{(\lambda)}-i\sin(q_{y}a)\exp(-iq_{z}^{(\lambda)}a)e_{y}^{(\lambda)}\right.\nonumber \\
 & \left.+\left(2-((\cos(q_{x}a)+\cos(q_{y}a))\exp(-iq_{z}^{(\lambda)}a))\right)e_{z}^{(\lambda)}\right]-\mathcal{M}(0,0)\left[3,3\right]e_{z}^{(\lambda)},\nonumber \\
C_{4\lambda} & =\mathcal{M}(0,0)\left[1,1\right]e_{x}^{(\lambda)},\nonumber \\
C_{5\lambda} & =\mathcal{M}(0,0)\left[2,2\right]e_{y}^{(\lambda)},\nonumber \\
C_{6\lambda} & =\mathcal{M}(0,0)\left[3,3\right]e_{z}^{(\lambda)}.\label{eq:coeff_reflected_wave}
\end{align}
The coefficients for the incident wave are ($\lambda=0$)
\begin{align}
C_{10} & =-K\left[\left(1-(\cos(q_{x}a)\exp(iq_{z}^{(0)}a))\right)e_{x}^{(0)}-i\sin(q_{x}a)\exp(iq_{z}^{(0)}a)e_{z}^{(0)}\right]\nonumber \\
 & +\mathcal{M}(0,0)\left[1,1\right]e_{x}^{(0)},\nonumber \\
C_{20} & =-K\left[\left(1-(\cos(q_{y}a)\exp(iq_{z}^{(0)}a))\right)e_{y}^{(0)}-i\sin(q_{y}a)\exp(iq_{z}^{(0)}a)e_{z}^{(0)}\right]\nonumber \\
 & +\mathcal{M}(0,0)\left[2,2\right]e_{y}^{(0)},\nonumber \\
C_{30} & =-K\left[-i\sin(q_{x}a)\exp(iq_{z}^{(0)}a)e_{x}^{(0)}-i\sin(q_{y}a)\exp(iq_{z}^{(0)}a)e_{y}^{(0)}\right.\nonumber \\
 & \left.+\left(2-((\cos(q_{x}a)+\cos(q_{y}a))\exp(iq_{z}^{(0)}a))\right)e_{z}^{(0)}\right]+\mathcal{M}(0,0)\left[3,3\right]e_{z}^{(0)},\nonumber \\
C_{40} & =\mathcal{M}(0,0)\left[1,1\right]e_{x}^{(0)},\nonumber \\
C_{50} & =\mathcal{M}(0,0)\left[2,2\right]e_{y}^{(0)},\nonumber \\
C_{60} & =\mathcal{M}(0,0)\left[3,3\right]e_{z}^{(0)}.\label{eq:coeff_incident_wave}
\end{align}
The coefficients $C_{p\lambda}$ for the three transmitted waves ($\lambda=4,5,6$)
are
\begin{align}
C_{1\lambda} & =\mathcal{M}({\bf q},q_{f\lambda})\left[1,1\right]e_{x}^{(\lambda)}+\mathcal{M}({\bf q},q_{f\lambda})\left[1,2\right]e_{y}^{(\lambda)}+\mathcal{M}({\bf q},q_{f\lambda})\left[1,3\right]e_{z}^{(\lambda)},\nonumber \\
C_{2\lambda} & =\mathcal{M}({\bf q},q_{f\lambda})\left[2,1\right]e_{x}^{(\lambda)}+\mathcal{M}({\bf q},q_{f\lambda})\left[2,2\right]e_{y}^{(\lambda)}+\mathcal{M}({\bf q},q_{f\lambda})\left[2,3\right]e_{z}^{(\lambda)},\nonumber \\
C_{3\lambda} & =\mathcal{M}({\bf q},q_{f\lambda})\left[3,1\right]e_{x}^{(\lambda)}+\mathcal{M}({\bf q},q_{f\lambda})\left[3,2\right]e_{y}^{(\lambda)}+\mathcal{M}({\bf q},q_{f\lambda})\left[3,3\right]e_{z}^{(\lambda)},\nonumber \\
C_{4\lambda} & =\left(\mathcal{U}({\bf q},q_{f\lambda})\left[1,1\right]-\mathcal{V}({\bf q},q_{f\lambda})\left[1,1\right]-\mathcal{M}({\bf q},q_{f\lambda})\left[1,1\right]\right)e_{x}^{(\lambda)}\nonumber \\
 & +\left(-\mathcal{V}({\bf q},q_{f\lambda})\left[1,2\right]-\mathcal{M}({\bf q},q_{f\lambda})\left[1,2\right]\right)e_{y}^{(\lambda)}\nonumber \\
 & +\left(-\mathcal{V}({\bf q},q_{f\lambda})\left[1,3\right]-\mathcal{M}({\bf q},q_{f\lambda})\left[1,3\right]\right)e_{z}^{(\lambda)},\nonumber \\
C_{5\lambda} & =\left(-\mathcal{V}({\bf q},q_{f\lambda})\left[2,1\right]-\mathcal{M}({\bf q},q_{f\lambda})\left[2,1\right]\right)e_{x}^{(\lambda)}\nonumber \\
 & +\left(\mathcal{U}({\bf q},q_{f\lambda})\left[2,2\right]-\mathcal{V}({\bf q},q_{f\lambda})\left[2,2\right]-\mathcal{M}({\bf q},q_{f\lambda})\left[2,2\right]\right)e_{y}^{(\lambda)}\nonumber \\
 & +\left(-\mathcal{V}({\bf q},q_{f\lambda})\left[2,3\right]-\mathcal{M}({\bf q},q_{f\lambda})\left[2,3\right]\right)e_{z}^{(\lambda)},\nonumber \\
C_{6\lambda} & =\left(-\mathcal{V}({\bf q},q_{f\lambda})\left[3,1\right]-\mathcal{M}({\bf q},q_{f\lambda})\left[3,1\right]\right)e_{x}^{(\lambda)}\nonumber \\
 & +\left(-\mathcal{V}({\bf q},q_{f\lambda})\left[3,2\right]-\mathcal{M}({\bf q},q_{f\lambda})\left[3,2\right]\right)e_{y}^{(\lambda)}\nonumber \\
 & +\left(\mathcal{U}({\bf q},q_{f\lambda})\left[3,3\right]-\mathcal{V}({\bf q},q_{f\lambda})\left[3,3\right]-\mathcal{M}({\bf q},q_{f\lambda})\left[3,3\right]\right)e_{z}^{(\lambda)}.\label{eq:coeff_transmitted_wave}
\end{align}

\end{document}